# Synergistic Graphene and PEDOT:PSS Layers on Silicon for Enhanced Efficiency and Stability in Schottky Junction Solar Cells

Kentaro Nakajo, Hojun Im*, Keiichiro Nishimura, Masahiro Teraoka, Masaya Kikuchi
Graduate School of Science and Technology, Hirosaki University, Hirosaki 036-8561, Japan

ABSTRACT

While graphene/n-type silicon (Gr/n-Si) and PEDOT:PSS/n-Si solar cells offer a cost-effective alternative to conventional photovoltaics, achieving simultaneous high power conversion efficiency (PCE) and long-term environmental stability remains a formidable challenge. Here, we report the synergistic enhancement of both performance and stability in a planar trilayer PEDOT:PSS/Gr/n-Si Schottky junction architecture. By systematically benchmarking this hybrid device against pristine Gr/n-Si and PEDOT:PSS/n-Si reference cells fabricated on untextured substrates, we isolate the fundamental optoelectronic and barrier functions of the carbon lattice. Comprehensive analyses reveal that the PEDOT:PSS layer simultaneously functions as an anti-reflection coating, a moderate p-type dopant for the underlying graphene, and a conductivity enhancer, collectively optimizing interfacial charge transport. Consequently, the trilayer device achieves a PCE of 8.52%, markedly outperforming the pristine Gr/n-Si reference cell (4.56%). Crucially, the atomic impermeability of the graphene sheet serves as a robust environmental barrier. While bare PEDOT:PSS/n-Si junctions retained only ~29% of their initial efficiency after 75 days of ambient exposure, graphene shields the active junction from moisture-induced degradation of the hygroscopic PEDOT:PSS overlayer. This enables the trilayer device to retain approximately 90% of its initial efficiency over the same 75-day period. The complementary roles of graphene as both a highly crystalline charge transport interface and an impermeable protective barrier establish a highly effective strategy for engineering high-efficiency, environmentally stable hybrid photovoltaics.

*Corresponding author. Email: hojun@hirosaki-u.ac.jp (H. Im)

## 1 INTRODUCTION

Novel Schottky junctions based on graphene and silicon have demonstrated outstanding optoelectronic characteristics over the last decade and have been recognized as a promising next-generation photovoltaic technology [1–6]. The strong performance of graphene/n-type silicon (Gr/n-Si) Schottky solar cells stems from the exceptional electrical and mechanical properties of graphene [6]. Its honeycomb crystal lattice, in which carbon atoms are tightly bonded through $sp^2$ hybridization, provides remarkable mechanical robustness and flexibility. Electrically, the unique band structure of graphene—where the valence and conduction bands converge at the Dirac point—enables extremely high Fermi velocities and carrier mobilities [7,8]. When interfaced with n-type silicon, graphene forms high-quality Schottky junctions governed by van der Waals interactions, typically exhibiting near-ideal diode behavior with ideality factors close to unity [9,10]. Despite these ideal interfacial properties, pristine Gr/n-Si Schottky junction solar cells generally deliver modest power conversion efficiencies (PCEs) of only 2–6% [9–13]. Substantial performance improvements have historically required complex, multi-step fabrication processes, such as applying volatile (e.g., $HNO_3$) or non-volatile (e.g., TFSA) p-dopants, depositing discrete anti-reflection coatings such as PMMA or $TiO_2$, and texturing the silicon surface [14–17], which have achieved PCEs up to ~17% [17]. However, combining these discrete optimization steps complicates device fabrication and frequently compromises long-term environmental stability. Consequently, there is a growing demand for multifunctional interlayers capable of simultaneously enhancing interfacial charge transport, optical absorption, and operational stability in a single step [18].

To fulfill this demand for a multifunctional interlayer, the conducting polymer PEDOT:PSS emerges as a highly promising candidate. Standalone PEDOT:PSS/n-type silicon (PP/n-Si) architectures have already demonstrated competitive efficiencies—reaching up to ~17% in optimized planar configurations through scalable, low-temperature solution processing [19–27]. In these cells, PEDOT:PSS has functioned simultaneously as a hole transport layer, a transparent conductive electrode, and an optical window [28–30]. However, a critical drawback of bare PP/n-Si cells is their severe moisture sensitivity which causes these devices to lose more than 60% of their initial PCE within 300 hours of ambient exposure due to the hygroscopic nature of the PSS [31,32].

In this work, we strategically incorporate a PEDOT:PSS overlayer into a Gr/n-Si Schottky junction to overcome the shortcomings of each individual architecture while leveraging their complementary strengths. To systematically isolate and quantify the fundamental optoelectronic and barrier roles of the carbon layer, we benchmarked the PEDOT:PSS/Gr/n-Si (PP/Gr/n-Si) architecture against pristine Gr/n-Si and PP/n-Si reference cells fabricated on untextured n-type silicon substrates. Through comprehensive optoelectronic, transient, and impedance analyses, we identify that the graphene layer plays a critical dual role. It provides a structurally conformal van der Waals contact to the n-type silicon substrate, and acts as an impermeable atomic-scale barrier that shields the active junction from moisture-induced degradation of the hygroscopic polymer overlayer. In addition, we demonstrate that the PEDOT:PSS layer functions simultaneously as an anti-reflection coating, a conductivity enhancer, and a moderate p-type dopant for the underlying graphene lattice. Consequently, the trilayer device achieves a PCE of 8.52%, nearly double that of the pristine Gr/n-Si reference (4.56%). Crucially, while bare PP/n-Si cells retained only ~29% of their initial efficiency after 75 days of ambient exposure, the trilayer maintained ~90% over the same period. These results establish that pairing a multifunctional polymer interlayer with the robust barrier properties and the conformal junction formation of graphene provides a highly effective strategy for engineering efficient and environmentally stable carbon-based hybrid photovoltaics.

## 2 EXPERIMENTAL

For systematic studies, we fabricated three types of solar cells with pristine Gr/n-Si, PP/n-Si, and PP/Gr/n-Si structures. The experimental procedure was divided into material preparation, device assembly, and performance characterization.

Graphene was synthesized on a 30 μm-thick copper foil via chemical vapor deposition (CVD). The reaction chamber was heated to 980 °C, and a gas mixture of Ar, $H_2$, and $CH_4$ was introduced for 45 minutes to facilitate graphene growth. For the transfer process, poly(methyl methacrylate) (PMMA, 1.0 wt% in anisole) was spin-coated onto the graphene surface at 3000 rpm as a support layer. The underlying copper foil was then etched using an aqueous iron(III) chloride ($FeCl_3$) solution, and the resulting PMMA/Gr film was thoroughly rinsed in deionized water to remove residual etchant.

An n-type silicon substrate (380 μm thick, 500 nm thermal oxide layer) was prepared through a series of cleaning and surface treatment steps. The substrate was first ultrasonically cleaned in acetone and isopropanol (IPA). Insulating polyvinyl tape with a ~2.5 mm diameter opening was then applied to define the active area. The substrate was subsequently treated with a buffered oxide etch (BOE) to remove the $SiO_2$ layer, followed by a piranha clean ($H_2SO_4$:$H_2O_2$) to eliminate organic contaminants, and a dilute hydrofluoric acid dip (DHF, 1 wt%) to strip any residual native oxide. Finally, the substrate was exposed to ambient air on a clean bench for two hours to allow the formation of a uniform native oxide layer for the passivation of the bare n-Si surface [16].

The PMMA/graphene film was carefully transferred onto the prepared silicon substrate by the wet-transfer method. To minimize damage to the graphene, the PMMA layer was removed via a multistep process comprising UV irradiation for 10 minutes, immersion in acetone at 50 °C, overnight storage in a vacuum desiccator, and thermal annealing at 420 °C for three hours under an Ar/$H_2$ atmosphere. For the PEDOT:PSS coating, a solution was prepared by mixing PEDOT:PSS with IPA, ethylene glycol, and Triton X-100, where the viscous Triton X-100 surfactant was included specifically to improve wetting and coating uniformity on the hydrophobic graphene surface. This solution was spin-coated over the active region at 1100 rpm for 240 seconds and subsequently baked on a hot plate at 150 °C for 10 minutes. For the electrical contacts, the back electrode was formed by applying a gallium–indium (GaIn) eutectic alloy to the scratched backside of the silicon substrate, while the front electrode was formed by applying silver paste around the active area.

Following fabrication, we carried out comprehensive structural, optical, and electrical characterization on the solar cells. Top-view images of the cells were obtained using an optical microscope, and cross-sectional images were acquired via scanning electron microscopy (SEM) using a JSM-7000F instrument (JEOL Ltd.). Raman spectroscopy, performed with a micro-Raman spectrometer (NRS-5100, JASCO) with a 532 nm excitation laser, confirmed the structural integrity of both the graphene and PEDOT:PSS layers. Photovoltaic performance was evaluated by measuring current density–voltage ($J$–$V$) characteristics under simulated solar illumination (AM 1.5G, 100 mW $cm^{-2}$) and in the dark, using a Keithley 2400 source measurement unit. Optical properties were assessed through external quantum efficiency (EQE) measurements using a monochromator (CS260, Oriel Ltd.), a lock-in amplifier (LI5640, NF Corp.), and a halogen lamp as the light source. Reflectance spectra were also collected over the 400–1100 nm wavelength range using a monochromator coupled with an integrating sphere detector. Carrier transport and recombination dynamics were further investigated via transient photocurrent (TPC) and transient photovoltage (TPV) measurements using a 639 nm laser diode with an output power of 10 mW (HL6358MG, Thorlabs) and an oscilloscope (MSO5354, Rigol). Impedance spectroscopy was performed in the dark over a frequency range of 10 Hz to 5 MHz using an LCR meter (ZM2376, NF Corp.).

3 RESULTS AND DISCUSSION

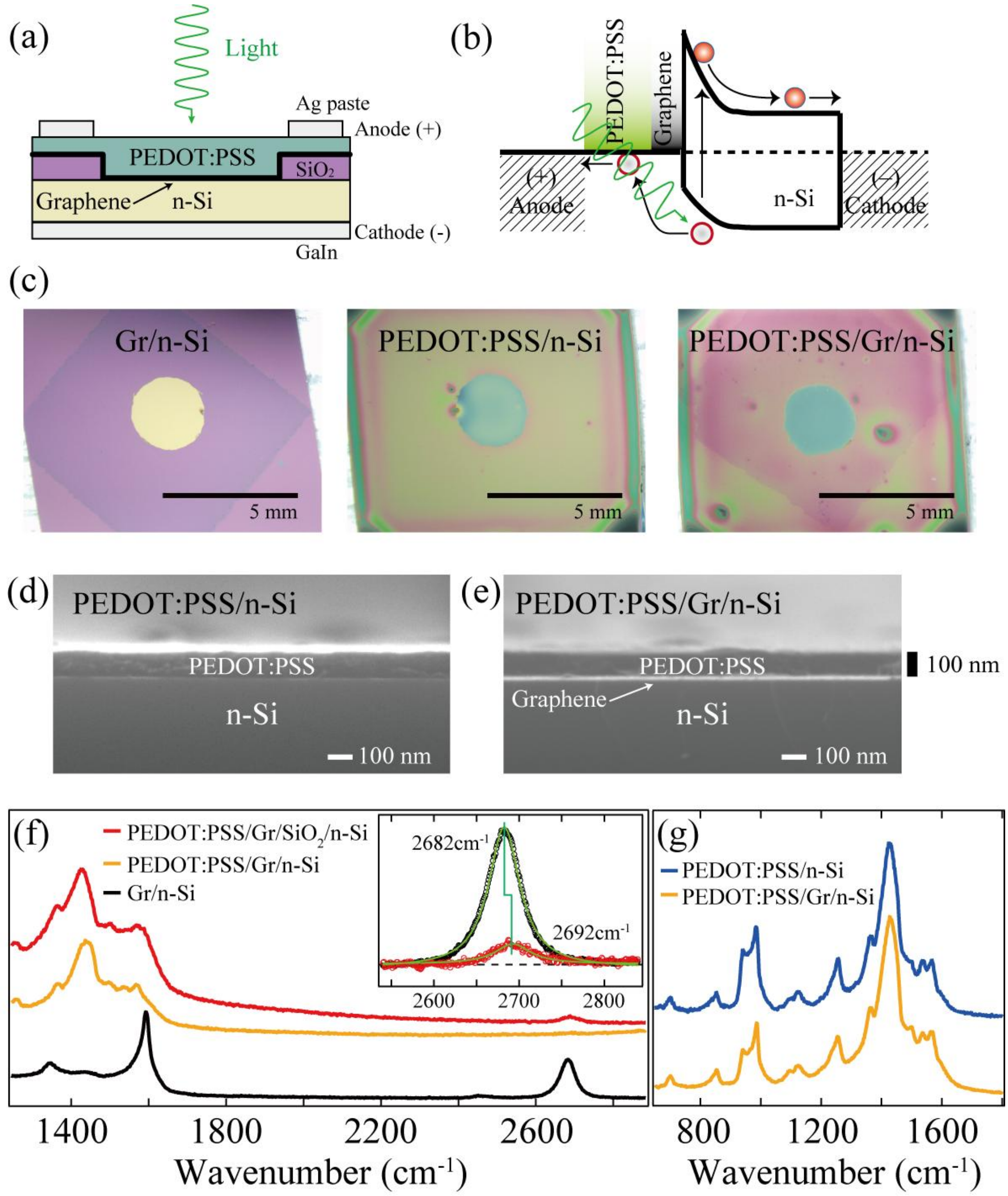

Figure 1. (a) Schematic illustration of the trilayer PEDOT:PSS/graphene/n-type silicon (PP/Gr/n-Si) Schottky junction solar cell and (b) the corresponding energy band diagram. (c) Optical images of the pristine Gr/n-Si (left), PP/n-Si (center), and PP/Gr/n-Si (right) solar cells. Cross-sectional SEM images of the (d) PP/n-Si and (e) PP/Gr/n-Si solar cells. Raman spectra of (f) Gr/n-Si, PP/Gr/n-Si, and PP/Gr/$SiO_2$/n-Si and (g) PP/n-Si and PP/Gr/n-Si. The inset of (f) shows the blueshift of the graphene 2D peak after deposition of the PEDOT:PSS overlayer.

Figure 1(a) schematically illustrates the trilayer PP/Gr/n-Si structure employed in this study. A Schottky junction is formed at the Gr/n-Si interface, onto which a PEDOT:PSS overlayer is subsequently deposited. The corresponding energy band diagram is presented in Fig. 1(b). The introduction of the PEDOT:PSS layer is expected to modulate the Schottky barrier height and charge transport across the junction, while simultaneously acting as an antireflection coating [22]. Figure 1(c) presents optical top-view images of the three fabricated solar cells: Gr/n-Si (left), PP/n-Si (center), and PP/Gr/n-Si (right). In the Gr/n-Si cell, a distinct purple-tinted square region is visible where the graphene was transferred onto the n-Si substrate, with the circular active area appearing grey at the center. In both the PP/n-Si and PP/Gr/n-Si cells, the active areas exhibit a pale green color, attributable to thin-film optical interference arising from the PEDOT:PSS coating. Figures 1(d) and 1(e) show cross-sectional SEM images of the PP/n-Si and PP/Gr/n-Si devices, respectively. We find that the spin-coated PEDOT:PSS layer has a thickness of approximately 100 nm, which has been considered as a proper thickness for antireflection [22]. In Fig. 1(e), the conformal graphene interlayer is clearly observed between the PEDOT:PSS overlayer and the n-Si substrate.

Raman spectroscopy was performed to investigate the crystallinity and structural integrity of the fabricated devices (Figs. 1(f) and 1(g)). The crystallinity is typically assessed by observing the G, 2D, and D bands. In high-quality SLG, the intensity ratio of the 2D to G bands ($I_{2D}/I_G$) is generally close to 2 and the D band exhibits low intensity. Figure S1 displays the Raman spectra of the graphene on $SiO_2$ substrate before and after thermal annealing. The spectra exhibit the characteristic G peak at ~1580 $cm^{-1}$, the 2D peak at ~2700 $cm^{-1}$, and the D peak at ~1350 $cm^{-1}$. Before thermal annealing, the Raman spectrum shows intensity ratios of $I_{2D}/I_G \sim 1.3$ and $I_D/I_G \sim 0.05$, confirming that the transferred graphene is highly crystalline SLG. After thermal annealing, the Raman spectrum retains the G and 2D peaks but develops a broad feature spanning 1300–1500 $cm^{-1}$. This band includes the D peak at ~1350 $cm^{-1}$, resulting in an increased $I_D/I_G \sim 0.5$ [33–36]. Such behavior was also observed in the Gr/n-Si cell used in this study as shown in Fig. 1(f). This broad spectral feature has been commonly observed in SLG subjected to thermal annealing for PMMA residue removal and is attributed to the carbonization of these residues [9,11,12,37].

In the PP/Gr/n-Si cell, the Raman spectra are dominated by PEDOT:PSS contributions due to the surface sensitivity of Raman measurements. It is well established that the Raman signal of graphene is significantly stronger on $SiO_2$ than on bare Si substrates [38]. To examine the shift in the 2D peak upon PEDOT:PSS deposition, the 2D peak of PP/Gr/$SiO_2$ on the same sample was also measured. As shown in the inset of Fig. 1(f), deposition of the PEDOT:PSS overlayer induces a pronounced blue shift (~10 $cm^{-1}$) of the graphene 2D peak, indicating a moderate p-type doping of the graphene layer [34,39]. In Fig. 1(g), the Raman spectra of the PP/n-Si and PP/Gr/n-Si cells show the C=C symmetric stretching mode of PEDOT at ~1450 $cm^{-1}$ and the sulfonic acid group stretching of PSS at ~1000 $cm^{-1}$, indicating the high structural quality of the PEDOT:PSS layer [40].

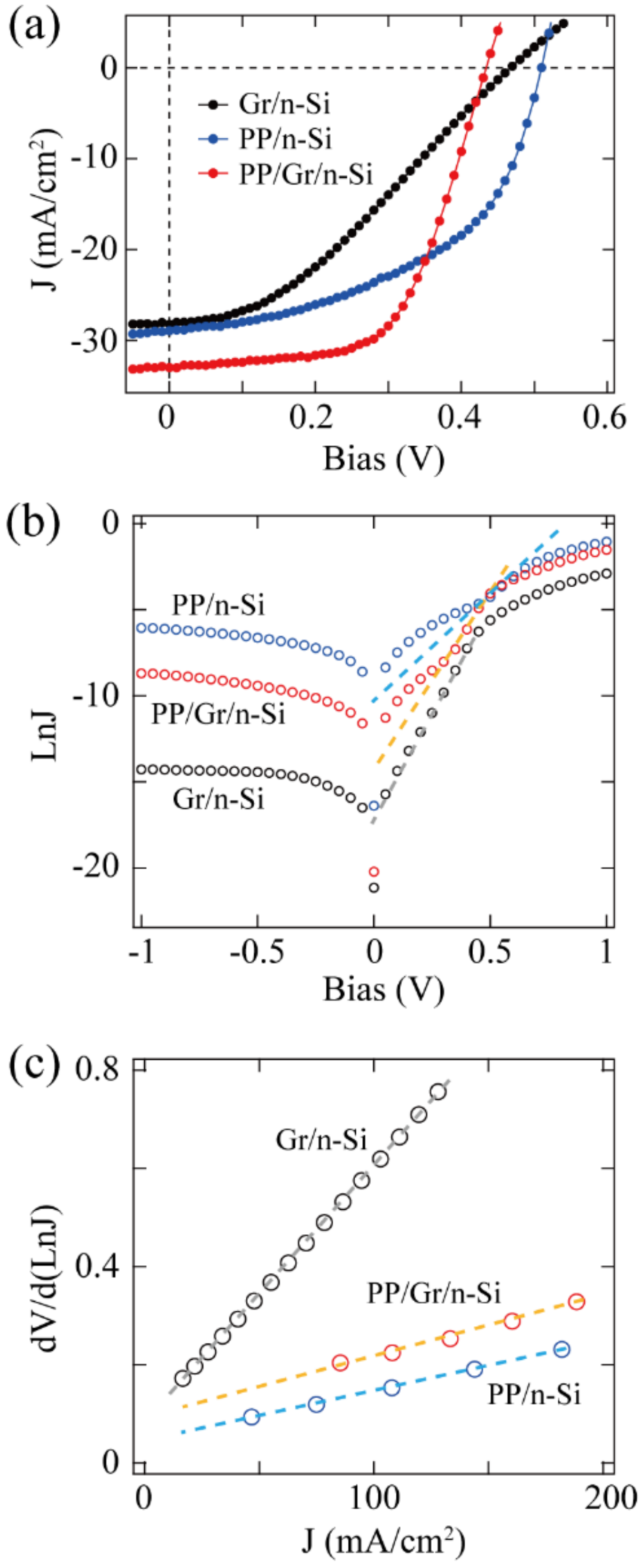

Figure 2. (a) J–V characteristics of the pristine Gr/n-Si, PP/n-Si, and PP/Gr/n-Si solar cells under illumination. (b) Dark lnJ–V plots used to extract the ideality factor and Schottky barrier height, and (c) Dark dV/d(ln J)–J plots used to extract the series resistance. Dashed lines serve as guides to the eye for curve fitting.

Figure 2(a) shows the J–V characteristics of the three device configurations under illumination. The pristine Gr/n-Si Schottky junction solar cell exhibits typical performance, with a PCE of 4.56% and a FF of 34.5%, consistent with previous studies (PCE ~ 2–6%) [4,9–13,15,16]. The $V_{OC}$ and $J_{SC}$ were measured to be 0.47 V and 28.2 mA $cm^{-2}$, respectively. By comparison, the PP/n-Si hybrid solar cell demonstrates a PCE of 7.46%, likewise in line with previously reported values for this device type [20,22,28–30]. Both the FF (50.5%) and $V_{OC}$ (0.51 V) are appreciably higher than those of the Gr/n-Si reference device, while the $J_{SC}$ (28.9 mA $cm^{-2}$) remains comparable. Notably, the trilayer PP/Gr/n-Si device achieves the highest performance among the three configurations, with the incorporation of the PEDOT:PSS overlayer boosting the PCE from 4.56% to 8.52% relative to the Gr/n-Si reference. This improvement is driven primarily by significant enhancements in both FF (59.5%) and $J_{SC}$ (33.0 mA $cm^{-2}$), although the $V_{OC}$ (0.43 V) decreases slightly compared to the Gr/n-Si reference.

The dark J–V curves were further analyzed to extract the ideality factor (n), Schottky barrier height ($\Phi_{SB}$), and series resistance ($R_S$) using the ln J–V and dV/d(ln J)–J plots as shown in Figs. 2(b) and 2(c), respectively. The pristine Gr/n-Si cell exhibits favorable diode parameters, with n = 1.74 and $\Phi_{SB}$ = 0.84 V, but a relatively high $R_S$ of 5.30 Ω $cm^2$. The PP/n-Si cell, by contrast, shows a large ideality factor (n = 3.11) and a reduced $\Phi_{SB}$ of 0.69 V, but benefits from a markedly lower $R_S$ of 1.04 Ω $cm^2$. The trilayer PP/Gr/n-Si cell yields the best overall diode parameters: n = 1.51, $\Phi_{SB}$ = 0.84 V, and $R_S$ = 1.30 Ω $cm^2$. This configuration effectively combines the advantages of both constituent device types, namely the favorable n and $\Phi_{SB}$ of the Gr/n-Si junction and the low $R_S$ of the PP/n-Si device, thereby synergistically enhancing photovoltaic performance. We note that the values of $\Phi_{SB}$ for the Gr/n-Si and PP/Gr/n-Si devices are almost identical. This indicates that the junction properties of the trilayer PP/Gr/n-Si device are primarily governed by the Gr/n-Si Schottky junction. Here, we briefly discuss the moderate p-doping of the graphene layer induced by the PEDOT:PSS overlayer, as evidenced by the Raman blue shift in Fig. 1(f). It is well established that p-doping graphene increases $\Phi_{SB}$ by lowering the Fermi level within the Dirac cone. However, this increase was not observed in the present trilayer PP/Gr/n-Si device. We tentatively attribute this to a compensation effect, where the expected increase in $\Phi_{SB}$ from p-doping is likely offset by a decrease in $\Phi_{SB}$ caused by the PEDOT:PSS deposition. Despite this limited impact on the barrier height, we expect the p-doping to remain beneficial for overall device performance, particularly for retaining photovoltaic efficiency. This warrants further investigation in future studies. The extracted photovoltaic parameters are summarized in Table 1.

Table 1. Photovoltaic parameters of three types of solar cells: pristine Gr/n-Si, PP/n-Si, and PP/Gr/n-Si.

| | $\eta$ [%] | $FF$ [%] | $V_{OC}$ [V] | $J_{SC}$ [mA/cm$^2$] | $n$ | $R_S$ [Ω cm$^2$] | $\Phi_{SB}$ [V] |
|---|---|---|---|---|---|---|---|
| Gr | 4.56 | 34.5 | 0.47 | 28.2 | 1.74 | 5.30 | 0.84 |
| PP | 7.46 | 50.5 | 0.51 | 28.9 | 3.11 | 1.04 | 0.69 |
| PP/Gr | 8.52 | 59.5 | 0.43 | 33.0 | 1.51 | 1.30 | 0.84 |

Impedance spectroscopy was also performed at zero bias in the dark, as shown in Fig. S2. From the Nyquist plots, the shunt resistance ($R_{SH}$) can be estimated from the diameter of the semicircle, which inversely correlates with leakage current. As shown in Fig. S2(a), the Gr/n-Si cell displays a non-closed semicircle, indicative of a very large $R_{SH}$ on the order of MΩ, consistent with previously reported values [12]. The PP/n-Si device, by contrast, exhibits an extremely low $R_{SH}$ of several kΩ, suggesting substantial leakage currents (Fig. S2(b)) [41]. The PP/Gr/n-Si device shows an intermediate $R_{SH}$ of approximately 0.1 MΩ (Fig. S2(c)), falling between those of the Gr/n-Si and PP/n-Si devices. These trends are in good agreement with the reverse current densities extracted from the J–V curves at −1 V in Fig. 2(a). Crucially, this enhancement in $R_{SH}$ in the PP/Gr/n-Si cell, compared to that of the PP/n-Si cell, provides evidence of the impermeability of the graphene sheet. This indicates that the structurally conformal graphene sheet serves as an effective moisture barrier. By physically decoupling the hygroscopic PEDOT:PSS overlayer from the underlying n-type silicon substrate, the graphene effectively passivates the surface against the parasitic leakage pathways typically induced by direct polymer contact.

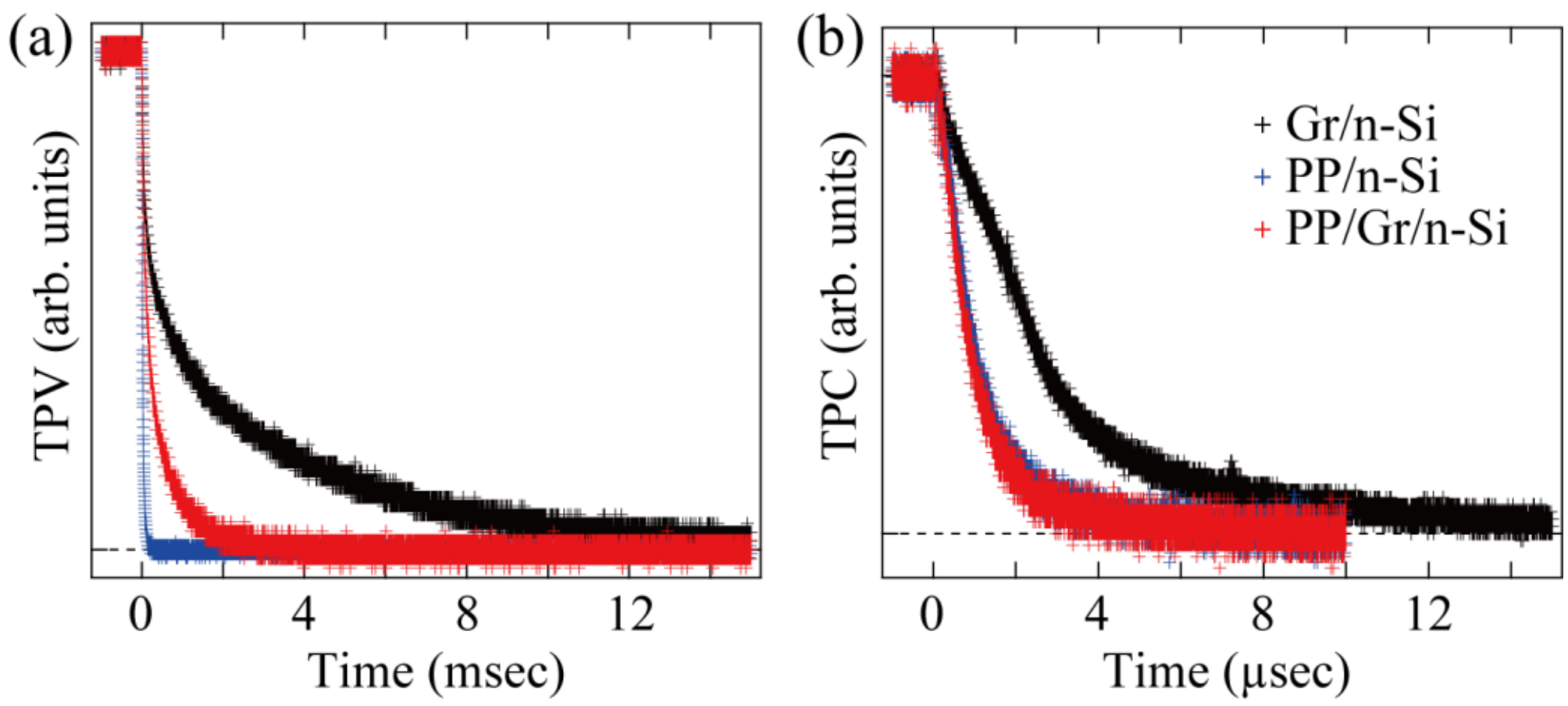

Figure 3. (a) Transient photocurrent (TPC) and (b) transient photovoltage (TPV) responses of three types of solar cells: pristine Gr/n-Si, PP/n-Si, and PP/Gr/n-Si.

Next, transient photovoltage (TPV) and transient photocurrent (TPC) measurements were carried out to investigate the recombination and charge transport dynamics of the three devices. The decay times, extracted from TPV and TPC, are interpreted as the recombination lifetime ($\tau_{rec}$) and the charge extraction time ($\tau_{ext}$), respectively [4,11,13,14]. As shown in Fig. 3(a), the TPV responses reveal that the Gr/n-Si cell exhibits the longest $\tau_{rec}$ of several milliseconds, while the PP/n-Si cell shows the shortest, on the order of tens of microseconds. Interestingly, the PP/Gr/n-Si cell displays an intermediate $\tau_{rec}$ of approximately 1 ms, falling between those of the other two devices. These trends are in agreement with the formation quality of the Schottky junction, revealed by photovoltaic parameters as shown in Fig. 2, including the ideality factor and Schottky barrier height.

As shown in Fig. 3(b), the TPC responses of all three devices exhibit rapid decay with $\tau_{ext}$ values on the order of several microseconds—approximately three orders of magnitude shorter than the corresponding TPV values—indicating efficient charge collection at the electrodes [9,11,13,14]. The PEDOT:PSS-based devices display faster charge extraction than the Gr/n-Si cell, consistent with their lower $R_S$ values, reflecting the high conductivity of PEDOT:PSS. Notably, the PP/Gr/n-Si device retains this fast extraction capability while simultaneously suppressing the recombination losses typically associated with direct PEDOT:PSS–silicon contact. This demonstrates the dual functional role of the graphene interlayer in this trilayer architecture.

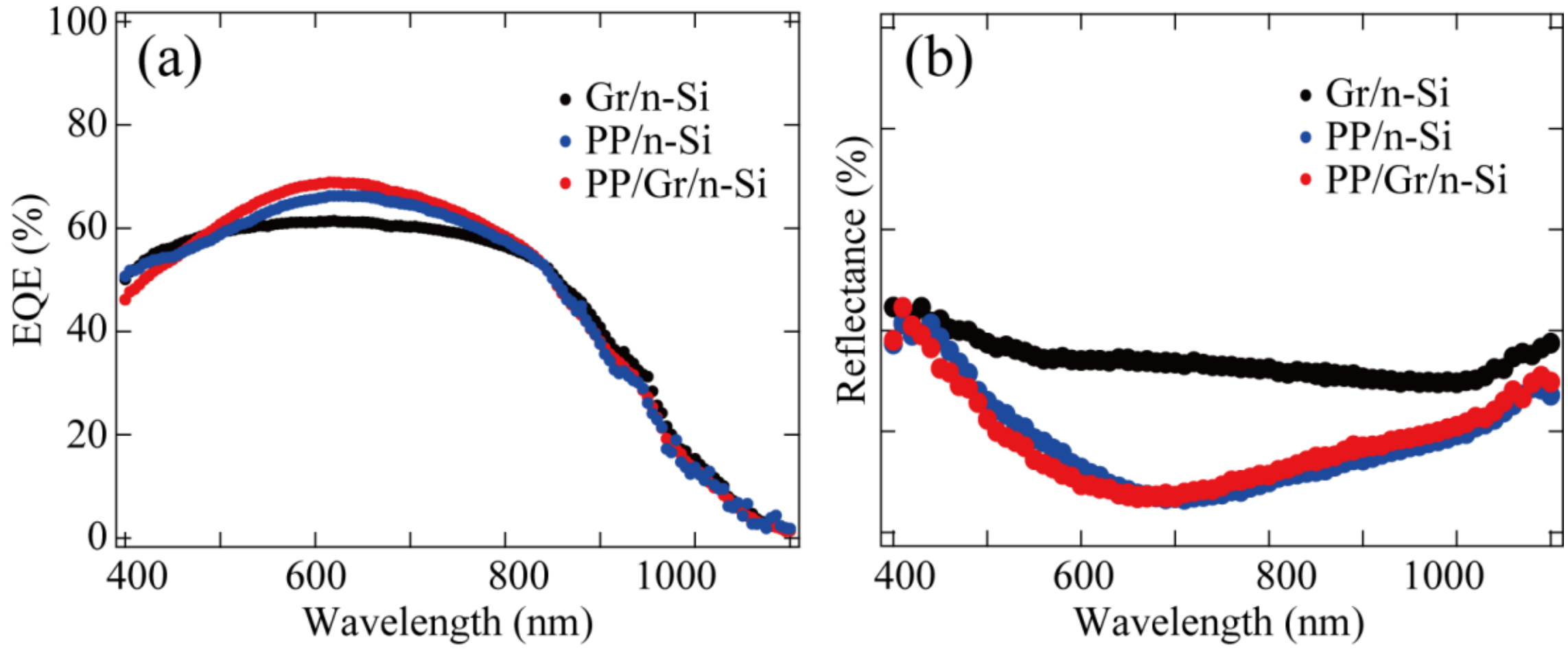

Figure 4. (a) External quantum efficiency (EQE) and (b) reflectance spectra of three types of solar cells: pristine Gr/n-Si, PP/n-Si, and PP/Gr/n-Si.

To investigate the photogenerated current as a function of wavelength across the visible to near-infrared range, EQE and reflectance measurements were performed from 400 to 1100 nm. Figure 4(a) shows the EQE spectra of the three device types. The Gr/n-Si cell displays a typical EQE spectrum characterized by a broad plateau of approximately 60% in the 400–800 nm range, consistent with earlier reports [11,13]. The PEDOT:PSS-based devices, by comparison, exhibit enhanced EQE with a peak of approximately 70% near 650 nm, which contributes to the improved $J_{SC}$ values observed in these devices.

Figure 4(b) displays the reflectance spectra of the three devices. The pristine Gr/n-Si cell exhibits high reflectance comparable to that of bare silicon, owing to the optical transparency of graphene. Upon incorporation of the PEDOT:PSS layer, the reflectance of both PEDOT:PSS-based devices drops markedly, reaching as low as approximately 5% near 700 nm. This substantial reduction confirms that PEDOT:PSS functions as a highly effective antireflection coating, consistent with the EQE enhancements observed near 700 nm in PEDOT:PSS-based devices in Fig. 4(a) [22,42].

It should be noted, however, that the $J_{SC}$ of the PP/n-Si cell is lower than that of the PP/Gr/n-Si cell and remains close to that of the Gr/n-Si cell despite the antireflection effect. This discrepancy is likely attributable to the higher recombination rate in the PP/n-Si device, as evidenced by the shorter $\tau_{rec}$ observed in the TPV measurements (Fig. 3(a)). This further suggests that the graphene interlayer promotes a higher-quality junction with the n-Si surface than direct PEDOT:PSS contact, in good agreement with the ideality factors extracted from the J–V analysis (Fig. 2(b)).

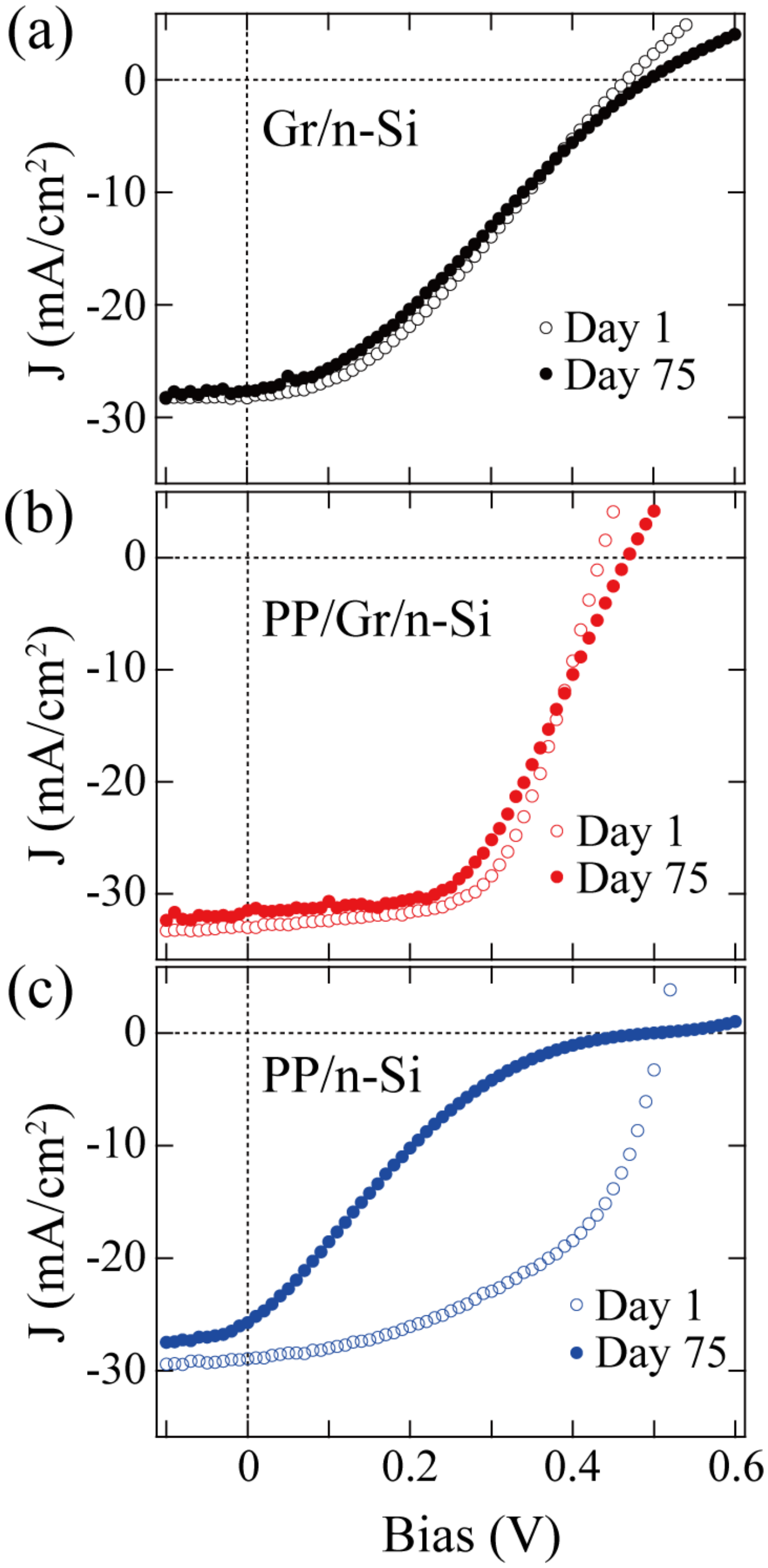

Figure 5. Long-term stability of the pristine Gr/n-Si, PP/n-Si, and trilayer PP/Gr/n-Si solar cells. The J-V characteristics under illumination are compared on day 1 and day 75 of ambient exposure.

Figure 5 presents the results of a 75-day stability test, highlighting a critical operational advantage of the trilayer architecture. All samples were stored under ambient atmospheric conditions at room temperature with a relative humidity of 20–30% throughout the testing period. The Gr/n-Si cell demonstrates excellent stability over the full testing period, retaining approximately 93% of its initial PCE and exhibiting a slight S-kink feature near the $V_{OC}$, consistent with previously reported values (Fig. 5(a)) [9–11,15]. A striking contrast is observed between the PP/n-Si and PP/Gr/n-Si cells. The PP/n-Si device exhibits significant PCE degradation after 75 days, retaining only approximately 29% of its initial value (Fig. 5(c)). This has been attributed primarily to the absorption of atmospheric moisture, which leads to increased $R_S$ and parasitic charge blocking at the polymer–silicon interface [31]. Remarkably, the highly efficient PP/Gr/n-Si device retains approximately 90% of its initial PCE over the same period (Fig. 5(b)), comparable to that of the Gr/n-Si cell (Fig. 5(a)). This robust stability is attributed to the highly stable Gr/n-Si Schottky junction within the PP/Gr/n-Si cell, and to the graphene interlayer, which effectively shields against moisture-induced degradation from the overlying PEDOT:PSS layer.

## 4 CONCLUSIONS

In conclusion, we have demonstrated a highly effective strategy for overcoming the efficiency–stability trade-off in carbon-based hybrid photovoltaics by engineering a planar trilayer PP/Gr/n-Si Schottky junction. Through the systematic integration of a multifunctional PEDOT:PSS overlayer, the hybrid device achieves a remarkable power conversion efficiency of 8.52%, substantially outperforming both the pristine Gr/n-Si (4.56%) and PP/n-Si (7.46%) reference cells. This enhancement is driven primarily by significant gains in short-circuit current density (33.0 mA $cm^{-2}$) and fill factor (59.5%).

Comprehensive optoelectronic characterization successfully decoupled the distinct contributions of the carbon lattice and the polymer overlayer. The PEDOT:PSS top layer acts as a highly effective antireflection coating, reducing surface reflectance to approximately 5% near 700 nm, while concurrently promoting rapid carrier transport with reduced series resistance. The conformal graphene interlayer establishes a high-quality van der Waals contact with the n-Si substrate, effectively suppressing the elevated interfacial recombination rates and parasitic leakage currents typically induced by direct polymer–silicon contact. Beyond its electronic role, the intrinsic atomic impermeability of the graphene sheet provides a robust environmental barrier, shielding the active junction from the moisture-induced degradation inherent to the hygroscopic PEDOT:PSS overlayer. As a result, the trilayer PP/Gr/n-Si cell retains approximately 90% of its initial efficiency after 75 days of ambient exposure. These results underscore the synergistic effects of the graphene and PEDOT:PSS layers in the Schottky junction photovoltaic devices, paving the way for high-performance, environmentally stable carbon-based optoelectronics.

SUPPLEMENTARY DATA

# Synergistic Graphene and PEDOT:PSS Layers on Silicon for Enhanced Efficiency and Stability in Schottky Junction Solar Cells

Kentaro Nakajo, Hojun Im*, Keiichiro Nishimura, Masahiro Teraoka, Masaya Kikuchi

Graduate School of Science and Technology, Hirosaki University, Hirosaki 036-8561, Japan

*Corresponding author. Email: hojun@hirosaki-u.ac.jp (H. Im)

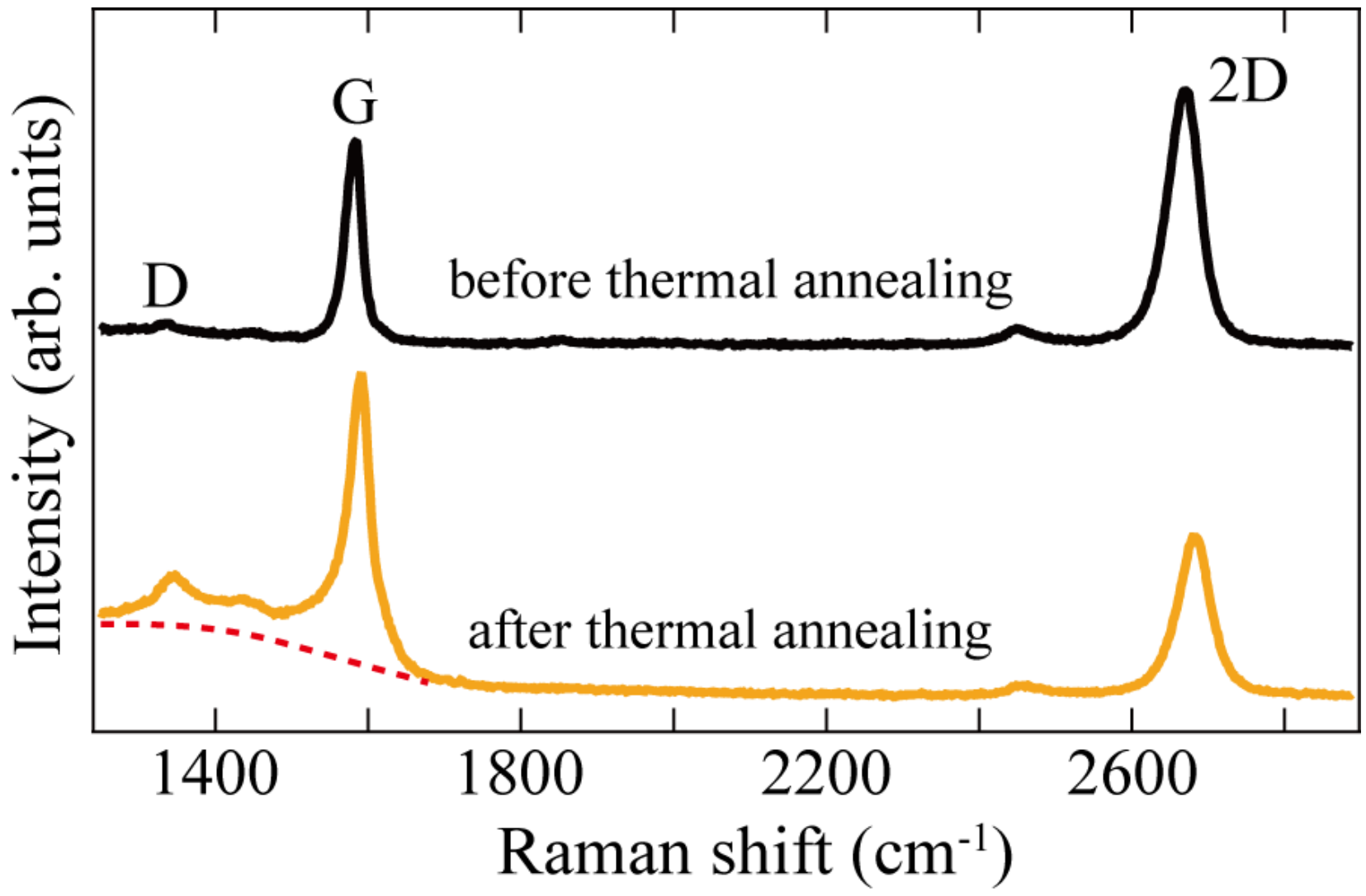

Figure S1. Raman spectra of graphene on $SiO_2$ before and after thermal annealing to remove PMMA residues. Before thermal annealing, the high $I_{2D}/I_G$ (~1.3) and low $I_D/I_G$ (~0.05) ratios indicate that the graphene is highly crystalline. After annealing, a broad peak appears between 1300 and 1600 $cm^{-1}$, originating from the carbonization of PMMA residues. The dashed red line is a guide to the eye for the broad peak.

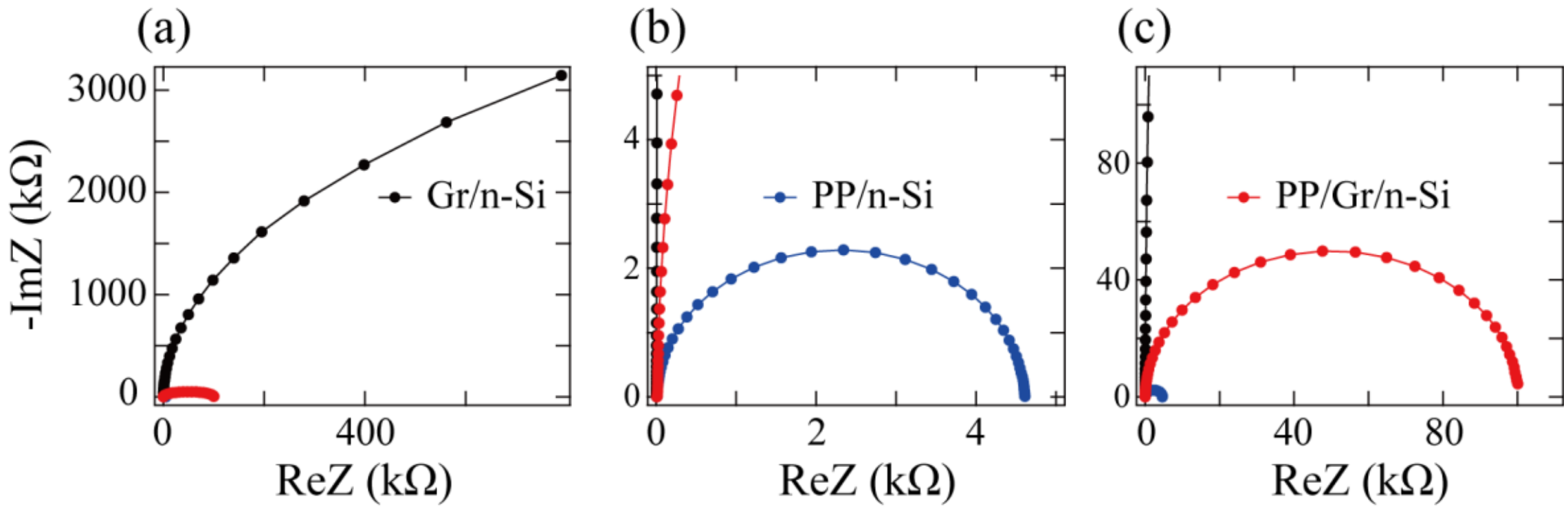

Figure S2. Nyquist plots of three fabricated solar cells: (a) pristine Gr/n-Si, (b) PP/n-Si, and (c) PP/Gr/n-Si, measured at zero bias in the dark. For direct comparison, the data for all three cells are presented together in each panel.

Table S1. Photovoltaic parameters of three types of solar cells: pristine Gr/n-Si, PP/n-Si, and PP/Gr/n-Si, measured after 75 days.

| | $\eta$ [%] | $FF$ [%] | $V_{OC}$ [V] | $J_{SC}$ [mA/cm$^2$] | $n$ | $R_S$ [Ω cm$^2$] | $\Phi_{SB}$ [V] |
|---|---|---|---|---|---|---|---|
| Gr | 4.24 | 31.1 | 0.49 | 27.6 | 1.51 | 11.96 | 0.90 |
| PP | 2.14 | 16.7 | 0.50 | 25.7 | 5.53 | 4.18 | 0.70 |
| PP/Gr | 7.65 | 52.0 | 0.48 | 31.5 | 1.55 | 4.08 | 0.84 |